\documentclass{jfm}
\usepackage{graphicx}
\usepackage{epstopdf, epsfig}
\usepackage{amsmath}
\usepackage{cite}
\usepackage[hidelinks]{hyperref}

\usepackage{todonotes}

\shortauthor{B. Boom, T. Truscott, F. E. Fish and E. Habtour}

\title{Tuning body shape and stiffness to mitigate water-entry forces}

\author{B. Boom\aff{1}
  \corresp{\email{bboom@uw.edu}},
  T. Truscott\aff{2}\corresp{\email{tadd.truscott@kaust.edu.sa}},
  F. E. Fish \aff{3}
 \and E. Habtour\aff{1} \corresp{\email{habtour@uw.edu}}}

\affiliation{\aff{1}Department of Aeronautics \& Astronautics, University of Washington,
Seattle, WA 98105, USA
\aff{2}Department of Mechanical Engineering, King Abdullah University of Science and Technology, Thuwal, 23955, Saudi Arabia
\aff{3}Department of Biology, West Chester University, West Chester, PA 19383, USA}

\begin{document}

\maketitle

\begin{abstract}

High-speed water entry of projectiles and diving systems induces high forces and jerk to the entering bodies due to the development of large hydrodynamic pressure. Previous research has shown separately that the peak forces can be reduced by improving the aerodynamic shape of the head (nose) or, recently, by introducing a spring element between the head and body. This study seeks to understand whether the aerodynamic shape or spring stiffness coupling is most important for force reduction by combining both in one study. The experiment combines the nose cone aerodynamics and spring stiffness with a rear body and examines the forces acting on the nose and body. Three parameters are varied: the nose angle, spring stiffness, and impact velocity. An unsteady semi-analytical formulation is developed to estimate the water entry forces and coupled body dynamics. We find that the peak force reduction due to the spring is highest when the slamming force is most significant, particularly at higher impact velocities and with blunter nose angles. The spring coupling enables periodic fluctuations between the kinetic and potential energy throughout the duration of impact, which can be tuned by varying the stiffness. These findings can allow engineers to control the dynamic response of water entry. 

\end{abstract}


\section{Introduction}

Large forces can be experienced by an object as it enters the water. Diving birds such as the Gannet (\textit{Morus bassanus}) can experience these forces on impact and have several features that reduce the impact forces: from their narrow beaks, and the structural complexity of their necks, to the pressure dissipation through their feathers \citep{weiss2014gaining, chang2016seabirds, bhar2019localized, sharker2019water, zimmerman2019nonlinear, zimmerman2020enhanced}. Understanding how these remarkable creatures mitigate the impact forces is both interesting and important for researchers and engineers attempting to reduce the forces of water impact for naval applications such as boat hull slamming \citep{abrate2011hull}, projectile water entry \citep{truscott2014water}, transformer drones \citep{zeng2022review,rockenbauer2021dipper, weisler2017testing}, and water plane float landing \citep{vonKarman1929impact}. 

Reducing these high impact forces has been the subject of concern in many engineering fields, as highlighted by \cite{truscott2014water}. The earliest measurements of the forces of water impact were made by \cite{thompson1928water, may1970review, grady1979hydroballistics, korobkin1988initial}. Researchers proposed several methods for reducing forces, such as using; surface treatments and features \citep{ARISTOFF_BUSH_2009, truscott2009water, shokri2022experimental, mehri2021water, benschop2017drag, koeltzsch2002flow, chen2014flow, chen2013biomimetic}, air jet \citep{elhimer2017influence}, water jet \citep{speirs2019water} and objects creating cavities in front of the projectile \citep{guo2020,lyu2021influence,rabbi2021impact}. From a nose geometry perspective, researchers have gained insights from diving birds into the relationship between beak shapes and functions, such as their ability to penetrate water surfaces at high velocities safely \citep{sharker2019water, weiss2014gaining, eliason2020morphological}. A wide range of nose geometries was investigated in great detail, for examples, spheres, ogives, cones, cusps and cups \citep{baldwin1971vertical,baldwin1975vertical,belden2023water,mcgehee1959water, thompson1965dynamic, li1967study, may1970review, qi2016investigation, sharker2019water,guzel2020reducing,xu2011numerical}.
A few studies focused on the effect of cone angle from both experimental \citep{baldwin1971vertical,may1970review} and analytical studies \citep{shiffman1951force,korobkin2006three,scolan2001three}. \cite{may1970review} measured the impact forces of cones and found nearly an order of magnitude difference between sharp ($C_d\sim 0.5$ for 22.5\textdegree\ half angle), to blunt shapes ($C_d\sim 6$ for 70\textdegree\ half angle) shapes. Most analytical models focused only on blunt cones, where blunt was defined as cones with half angles between 70\textdegree\ and 84\textdegree\ \citep{malleron2007some}. Furthermore, sharper cones experience the impact over more extended time periods than blunt cones. In fact, \cite{sharker2019water} found that the cone angle had a significant effect on jerk (the derivative of acceleration) and suggested that the diving birds' head shape were designed to reduce jerk. Indeed, jerk is an important marker to asses passenger injury risk in crash testing and other impact events, for example in automotive, aircraft, train and marine vehicles \citep{hayati2020jerk, eager2016beyond}.

\begin{figure}
    \centering
\includegraphics[width=\linewidth]{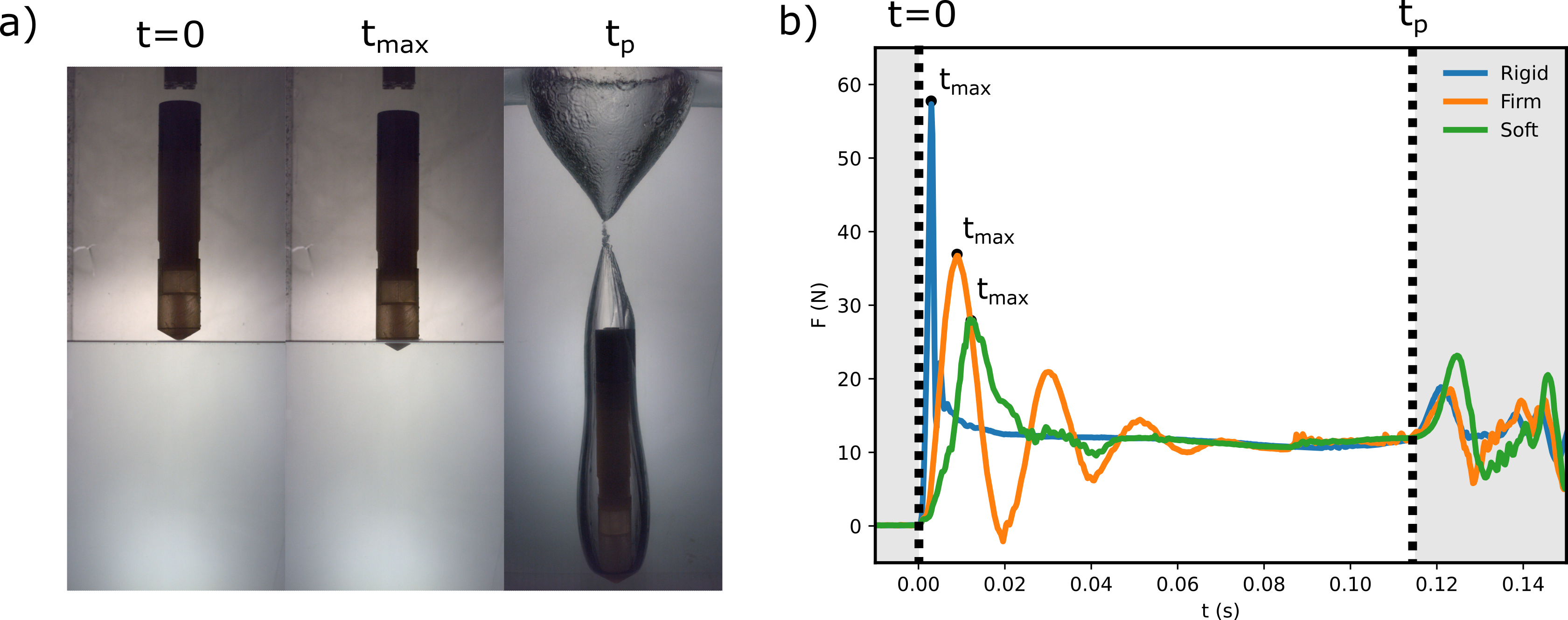}
    \caption{a) Highlights three important events during impact: i) $t=0$ initial impact; ii) $t_{max}$ time of maximal force; and iii) $t_p$ Pinch-off, the point where the cavity collapses. b) Shows the evolution of the forces acting on the body (back of the projectile). The time stamps of the three impact events are highlighted in the figure for a rigid projectile rigid, and with coupling springs (stiffness constants= 7.8 and 1.7 N/mm for firm and soft springs, respectively).}
    \label{fig:IntroFigure}
\end{figure}

The Gannet design features can potentially provide some invaluable cues into force reduction methodologies. One extraordinary feature of these birds is the neck design. Computed Tomography (CT) scans generated by \cite{chang2016seabirds} showed that the Gannet neck is an S-shape and presumably acts as a spring in compression along the midline contrary to previous assumptions. Using linear Euler–Bernoulli beam theory, \cite{chang2016seabirds} modeled the neck as a single beam, as a preliminary attempt, to estimate buckling due to impact. The next logical step is to represent the neck as an array of springs that couple the conical head to the body. Our proposed hypothesis in this study is that the compliance (e.g., spring) of the neck can reduce the slamming forces due to water entry. This hypothesis is supported by previous studies suggesting that reducing impact forces can be realized when an ogive nose is coupled to a body mass with a spring \citep{wu2020water, boom2023water, antolik2023slamming}. 

Only limited studies examined the water impact of a two-body system connected with a spring \citep{miller1952hydrodynamic, carcaterra2000prediction, wu2020water, antolik2023slamming, boom2023water}. \cite{miller1952hydrodynamic} modeled hydrodynamic impact due to seaplane landings where the aircraft structure was expressed as a linear spring. \cite{carcaterra2000prediction} they developed a nonlinear fluid-solid interactions model to predict peak forces for a two-mass wedge system impacting water at hypersonic speed. More recently, \cite{wu2020water, antolik2023slamming, boom2023water} studied the two-body sprung system with an ogive cone shape. \cite{wu2020water} demonstrated experimentally that the impact force decreases with the introduction of spring but did not provide an analytical relationship between the spring stiffness impact forces. \cite{antolik2023slamming} looked at multiple nose radii and multiple spring stiffnesses during water impact and provided a two-coupled model that showed both force reduction and increases on the body under certain conditions with the addition of a spring. \cite{boom2023water} compared experimentally the impact performance of a sprung system with ogive and 60\textdegree\ nose shapes and found both the shape and spring can reduce the maximal acceleration and jerk. 

There are two highly utilized dynamic modeling approaches for estimating the impact forces of water entry developed by \cite{Wagner1932} and \cite{vonKarman1929impact}. The von Wagner approach determines the reaction forces of the body by integrating the pressure across the wetted surfaces. The major challenge with this approach is that the potential flow must be known. To overcome this challenge, most of the reported studies simplify Wagner's approach by assuming an expanding flat plate and constant impact velocity \citep{abrate2011hull}. The von Karman approach assumes that all forces can be neglected except the inertia. This results in a simplified solution that only considers the inertia of the projectile mass and estimated added mass (approximation in time, based on a circle with the radius of the wetted diameter), but neglects buoyancy, surface tension, etc. There were several studies that combined \citep{Wagner1932} and \citep{vonKarman1929impact} approaches. For example, \cite{shiffman1951force} combined the von Karman and Wagner expanded plate approaches to analytically find the added mass by using a linearization and iteration method suggested by \cite{hillman1946vertical}. \cite{antolik2023slamming} modified the von Karman approach by using the velocity at impact to calculate the impact forces in time and thus neglecting the inertial term associated with the added mass but keeping the change in added mass associated with the drag term.

The nonlinear equations of motion developed in our study treated the added mass and body velocity during impact as time-varying parameters. Our approach was built on the works of \cite{carcaterra2000prediction, ARISTOFF_BUSH_2009, baldwin1971vertical}, where we included the contribution of steady state drag, the nose cone mass, body buoyancy, and the time evolution of the added mass as an exponential decay term. We also point out that the peak force does not occur at full cone submergence, as suggested by most analytical models and also found in wedge impact \citep{vincent2018dynamics}. The model can be utilized to estimate the response of projectiles with heavy and blunt noses, as well as sharp and light noses. The model was represented in the common form seen in the fields of aeroelasticity and nonlinear structural dynamics \citep{bisplinghoff2013principles, habtour2022highly}, which permitted a straightforward assessment of the effects of the hydrodynamic forces on the body dynamics during water entry, and the nonlinear contributions of inertia, hydrodynamic stiffness, and hydrodynamic damping. This modeling approach also estimates the energy transmission and dissipation during impact. 

In this study, we compared the influence of two design features, nose-cone geometry (angle) and spring stiffness, on the slamming forces experienced by a body during water entry. Here, we investigated nose angles ranging from 10\textdegree\ to 80\textdegree\ and two spring stiffnesses 1.74 and 8.16 N/mm.  The aim was to answer two fundamental questions. First, which design feature was most influential in reducing the slamming forces? Second, how the interplay between nose-cone geometry (angle) and spring stiffness could be exploited to manipulate the flow of the impact energy? \autoref{fig:IntroFigure} represents examples of three experimental prototypes that consist of conical shape noise bodies with no spring (rigid) and with firm, and soft springs. In \autoref{fig:IntroFigure}-a, we show an example of a nose-spring-body prototype experiencing three critical events during an impact: (i) initial impact at $t=0$; (ii) followed by maximum deceleration of the body at $t_{max}$; and (iii) pinch-off at $t_p$, the point where the cavity collapses. The corresponding acceleration for each stage is shown in \autoref{fig:IntroFigure}-b. We show that adding a spring did not alter the efficiency of water entry when compared to a rigid body but rather delayed the dissipation of energy. The delay was achieved by temporal potential storage of some of the body's kinetic energy in the spring, which was released at a later time. This effect was most significant for blunt nose angles and negligible for sharp nose angles.

The paper is structured as follows. \autoref{sec:ExpSetup} details the experimental setup and data acquisition. The effect of only the nose geometry on the impact force is provided in \autoref{sec:shape}. The combined effect of spring stiffness and nose shape is detailed in \autoref{sec:stiffenss}. Additional analyses of the energy flow, water entry efficiency and jerk are discussed in \autoref{sec:Results}.

\section{Experimental methods}\label{sec:ExpSetup}

\subsection{Experimental Design}
The bodies were dropped at different heights, $H$, using an electromagnet controller switch. The dropped bodies were accelerated by gravity, as shown in  \autoref{fig:ExpermintalSetup} a). The height ranged between 0.25 and 1.5 m to reach impact velocity $U_0 \approx \sqrt{2gH} \approx 2.2$ to $5.4 \text{m/s}^{-1}$, respectively. The bodies were dropped into a round tank with internal dimensions of 895 $\times$ 514 mm ($H \times D_{tank}$). Wall effects were neglected since the ratio of the dropped body diameter, d, and the tank diameter, $D_{tank}$ was $d/D_{\text{tank}} \approx 1/10$ \citep{guo2020}.

\subsection{Prototype design}
\begin{figure}
    \centering
    \includegraphics[width=0.8\linewidth]{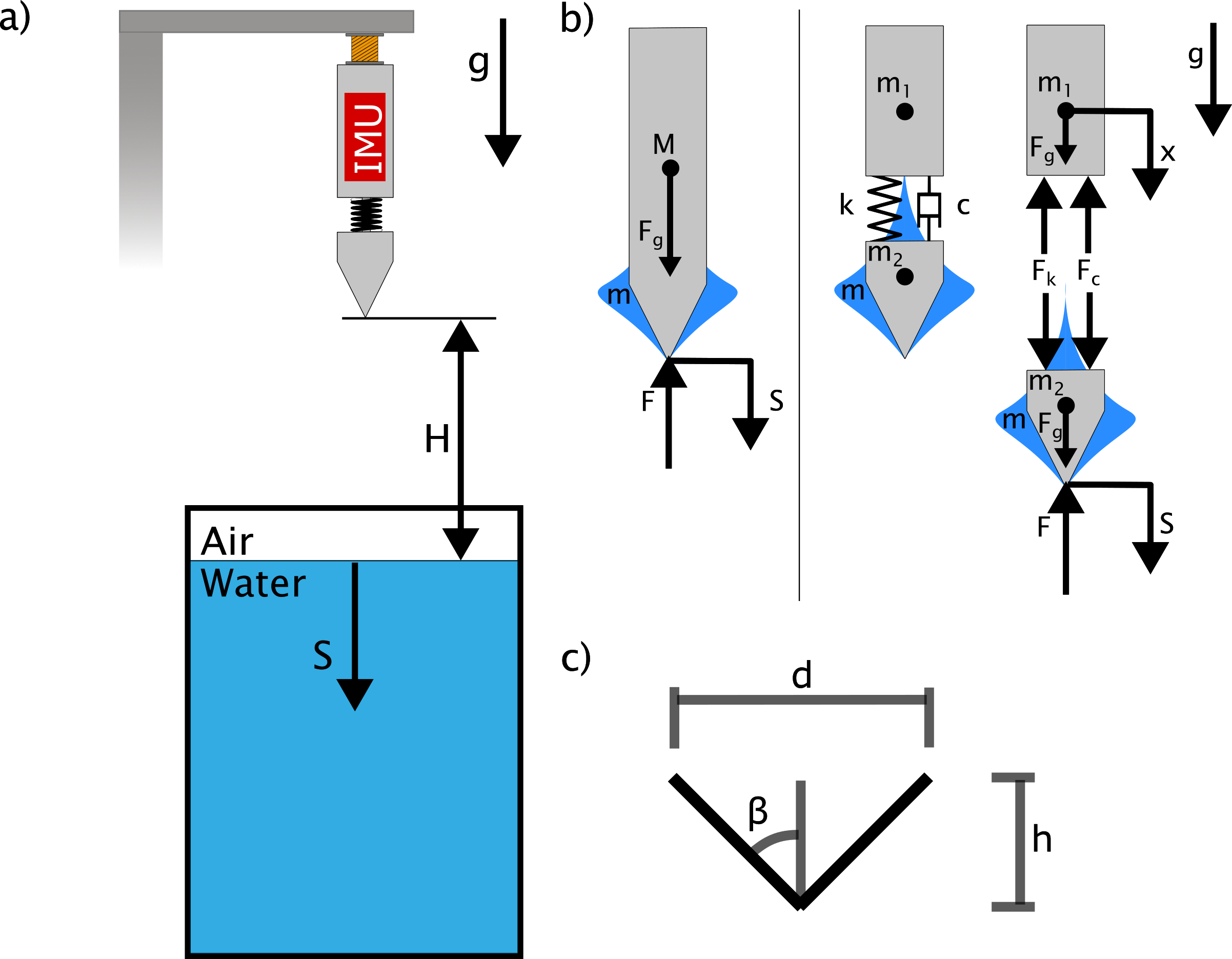}
    \caption{a) Schematic of the experimental setup, where $H$ and $S$ are the drop height and depth traveled by the projectile. The drop release is controlled using an electromagnetic device shown in orange. The projectile is accelerated by gravity. The untethered inertial measurement unit (IMU) is housed in the rear body and provides collection and storage of acceleration measurements. b) Free body diagram of rigid and sprung projectiles. c) cone diameter, half cone angle, and cone height are $d$, $\beta$, and $h$, respectively. The range of $\beta$ is 10\textdegree\ to 80\textdegree.}
    \label{fig:ExpermintalSetup}
\end{figure}

Each test projectile consisted of a conical nose (head) and a body connected with a spring or a rigid structure, see free body diagram in \autoref{fig:ExpermintalSetup} b). The head and body were 3D printed using Polylactic acid (PLA), and Formlabs tough 2000 resin, respectively. The Formlabs resin was utilized to ensure waterproof internal housing. The head geometry is depicted in \autoref{fig:ExpermintalSetup} c). The diameter, d, was kept constant at 52 mm. Ten different half cone angles, $\beta$, between 10\textdegree\ and 80\textdegree\ were tested, see \autoref{tab:Cones}. The nose dimensions and mass for each tested head are provided in \autoref{tab:Cones}. The rear bodies were connected rigidly or with either firm or soft springs with stiffness constants $k_S = 1.74 \text{N/mm}$ and $k_F = 8.16$ N/mm, respectively. To ensure that the cone moved only along the body's vertical axis, the nose motion was constrained with a linear bearing, which had minimal stiction but exhibited small viscous damping, as detailed in \autoref{sec:stiffenss}. The bearing housing was connected to the cone; as such, the nose and body masses can be computed as $m_2 = m_{cone}+m_{\text{linear bearing}}$, and $m_1 = m_{body}+m_{sping}$, respectively. The mass of each component in the projectile is provided in \autoref{tab:parts} and \autoref{tab:Cones}. The body contained an inertial measurement unit (IMU) and a steel plate at the back to make it possible to be dropped with the electromagnetic release mechanism.

\subsection{Experimental measurements}
Acceleration measurements were collected using an accelerometer sensor embedded in the projectile's body. The IMU maker was enDAQ (Model: S2000D40), which had a 4000 Hz sampling frequency, 0.00008 g resolution, and less than 0.01 gRMS noise level.

It was necessary to estimate four non-dimensional numbers, the Weber number ($We$), Froude number ($Fr$), Reynolds number ($Re$), and Bond number ($Bo$) to discern the dominant hydro effects that must be included in the analysis. They are defined as $We = \rho dU_0^2/\sigma$, $Fr = U_0/\sqrt{gd}$, $Re = U_0d/\nu$ and, $Bo = We/Fr^2 = \rho g d^2/\sigma$, where $\rho = 1000$ $kg m^{-3}$, $\sigma = 70\times 10^{-3}$ $N m^{-1}$, $g = 9.81$ $ms^{-2}$, $\nu = 10^{-6}$ $m^2 s^{-1}$, $U_0$, and $d$ are the water density, surface tension of the air-water interface, gravitational constant and kinematic viscosity, water entry velocity and characteristic diameter, respectively. In this study, the estimated four non-dimensional numbers were $We= 3.6 \times 10^{3}$ to $22 \times 10^{3}$, $Fr=3.08$ to $7.56$, $Re=1.1 \times 10^{5}$ to $2.8\times 10^{5}$, and $Bo= 378.9$. The high Weber number meant that the inertia of the displaced water was much more significant than the surface tension forces, which could be neglected. The hydrostatic and viscous forces could not be neglected due to low $Fr$ and transition $Re$, respectively. The Bond number compares the strength of the surface tension to the gravitational forces. The high Bond number confirmed that the system was unaffected by surface tension surface.

\begin{table}
\def~{\hphantom{0}}
\begin{center}
    \parbox{.45\linewidth}{
    \centering
        \begin{tabular}{ccc}
            Half angle $\beta$, \textdegree \quad \quad & $h_{cone}$, mm & Mass, g \\ \hline
            10 & 147.4 & 45.8\\
            12 & 122.3 & 43.6\\
            25 & 55.8  & 31.2\\
            30 & 45.0 & 28.3\\
            40 & 31.0 & 32.1\\
            45 & 26.0 & 30.1\\
            50 & 21.8 & 30.7\\
            60 & 15.0 & 30.0\\
            70 & 9.5 & 28.0\\
            80 & 4.6 & 25.7
        \end{tabular}
        \caption{Cone dimensions and masses}
        \label{tab:Cones}
    }
    \hfill
    \parbox{.45\linewidth}{
    \centering
        \begin{tabular}{lcc}
        Part & Mass, g \\ \hline
        Body & 522.8\\
        Linear bearing & 34.28\\
        Rigid piece & 2.5\\
        Firm spring & 4.0 \\
        Soft spring & 1.5\\
        &\\
        &\\
        &\\
        &\\
        &
        \end{tabular}
        \caption{Component masses}
        \label{tab:parts}
}
\end{center}
\end{table}

\section{Cone shape effect on water-entry}\label{sec:shape}

This section details the development of an unsteady state semi-analytical model for estimating the water impact forces for different nose angles. The objective of the model is to describe in isolation the effect of the nose angle on the spatial and temporal evolution of the hydrodynamic forces starting from the initiation of water impact to the collapse of the air cavity. The modeling results were validated experimentally for ten different cone half angles ranging from 10\textdegree\ to 80\textdegree, see \autoref{tab:Cones}. The model considers the effect of buoyancy and velocity change throughout the impact to provide accurate predictions of the added mass coefficients and, thus, the reaction forces.

\subsection{Estimating impact forces} \label{sec:forceModel}
\begin{figure}
    \centering
    \includegraphics[width = 0.6\linewidth]{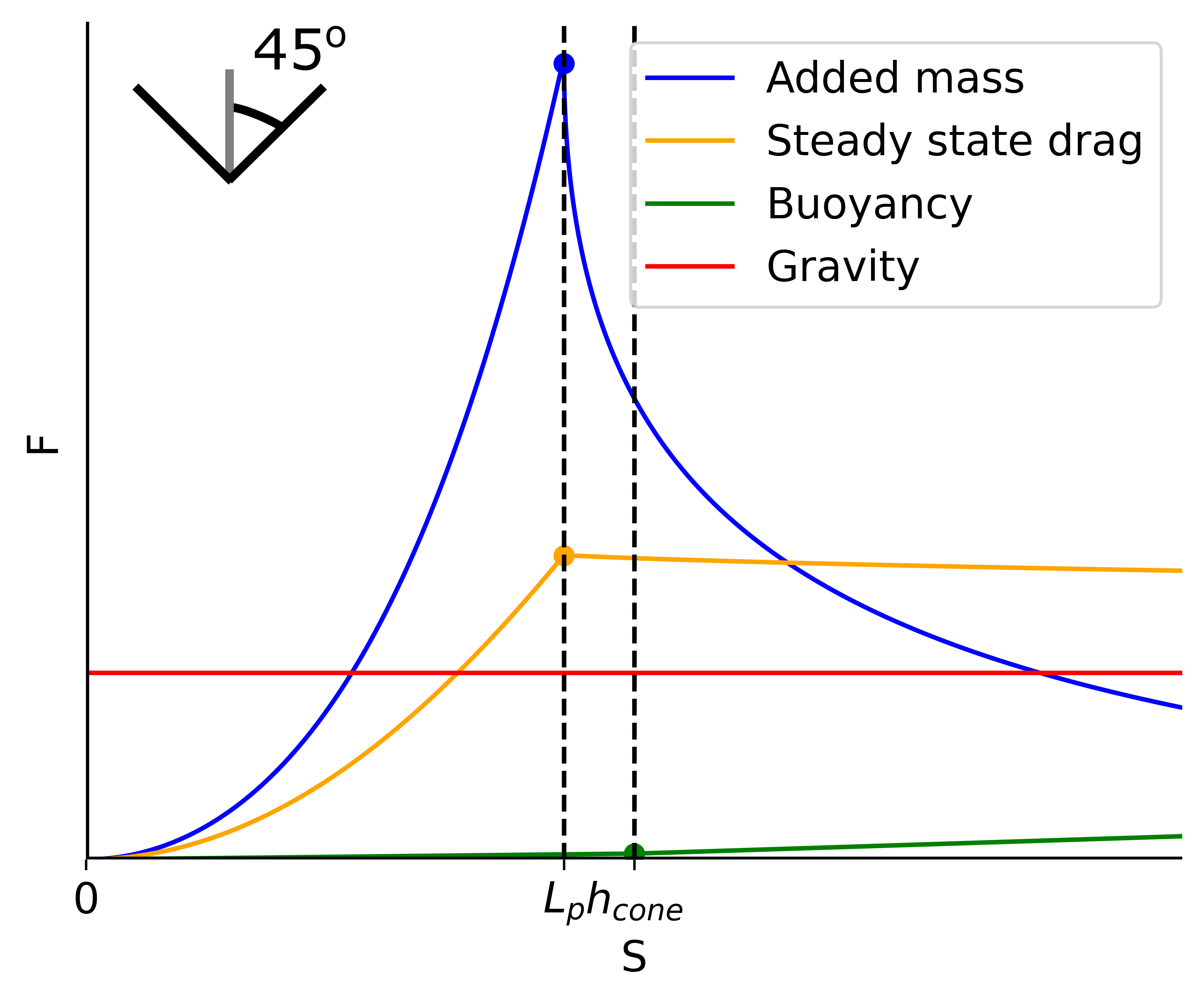}
    \caption{Breakdown of forces during impact modified from \cite{baldwin1971vertical}. The vertical lines depicted indicate instances where the depth $S$ is equal to $L_p$ and $h$, with $L_p$ representing the cone depth at peak acceleration and $h$ indicating when the cone is fully submerged.}
    \label{fig:ImpactForces}
\end{figure}

The model in this study includes the added mass effect, total drag, gravitational acceleration, and buoyancy, but neglects surface tension effects due to a high bond number, see \autoref{sec:ExpSetup}. As shown in \autoref{fig:ImpactForces}, the added mass and steady state drag are divided into two spatial events. 
The first event is modeled for $0<S< L_p$, which is from the point of initial impact($t=0$) until the peak force is reached($t_{max}$). The Second event in the model is for $S\geq L_p$, beyond the peak force until the cavity collapses($t_p$). It can also be noted that the buoyancy is modeled with two splits one at full cone submersion, $S = h_{cone}$, and one at full body submersion, $S = h_{cone}+h_{body}$. 

\subsubsection{Equation of motion for unsprung projectile}
From the free-body diagram in \autoref{fig:ImpactForces}-b, the equation of motion for both periods can be derived from equilibrium, as follows \cite{abrate2011hull};

\begin{equation} \label{eq:eom}
    \ddot{S}(M+m) -\frac{dm}{dS}\dot{S}^2 =  - F_g + F_b + F_{Cds}
\end{equation}

where, $\dot{S}$ is the velocity of $M$ and $m$, which are the mass of the body, and added mass, respectively. The gravity, buoyancy and steady state drag forces are $F_g$, $F_b$, and $F_{Cds}$, respectively. The surface tension effects are neglected due to the high Bond number, as stated in \autoref{sec:ExpSetup}. The first two terms in \autoref{eq:eom} are the inertial forces of $M$ and $m$ due to their change in their acceleration, $\ddot{S}$. The third term is the unsteady force of the added mass as it changes incrementally with depth; $dm/dS$. This force is quadratically dependent on the velocity; acting as a drag term. For convenience, \autoref{eq:eom} is rewritten to combine the unsteady force with steady drag force into $F_{Cd}$, since both are related to the square of the velocity, to get the full dynamics as:
\begin{equation}\label{eq:eomcd}
    (M+m) \ddot{S}- F_b - F_{Cd}  = - F_g
\end{equation}
where,
\begin{equation} \label{eq:HydroF}
    F_g = Mg, \quad F_{Cd} = \frac{\rho}{2} C_d(S) A_b(S) \dot{S}^2, 
    \quad F_b = \rho g V_s(S) 
\end{equation} 
\begin{align} \label{eq:C_d}
    C_d(S) &= C_{du}+C_{ds} = \frac{2}{\rho A_b(S)}\frac{dm}{dS}+C_{ds}\\
    V_s(S) &= \pi \tan{\beta}^2 \begin{cases}
          \frac{1}{3} S^3  & 0\leq S< h_{cone}\\
          S h_{cone}^2-\frac{2}{3} h_{cone}^3  & h_{cone}\leq S< h_{cone}+h_{body}\\
          h_{body}h_{cone}^2-\frac{2}{3} h_{cone}^3  & S\geq h_{cone}+h_{body}
    \end{cases}\label{eq:buoyancy}
\end{align}

where, $A_b(S)$ is the nose projected surface as a function of the depth $S$. The unsteady and steady coefficients of drag are $C_{du}$, and $C_{ds}$, respectively. It is important to point out that unlike previous studies $C_d(S)$ cannot assumed to be constant since it is a function of the change in the added mass and projected area with respect to the nose penetration depth, (\autoref{eq:C_d}). The buoyancy, defined above in \autoref{eq:HydroF}, is proportional to the length portion of the submerged volume of the projectile. Therefore, it is necessary to capture the submerged volume, $V_s(S)$, as a function of depth, which can be summarized into three stages, see \autoref{eq:buoyancy}. The first stage is $0\leq S <h_{cone}$, when the nose starts to submerge until penetration depth reaches $h_{cone}$. Note that due to the nose shape the buoyancy is cubic in the penetration depth. The second stage is when the cylindrical body starts to submerge, $h_{cone}\leq S< h_{cone}+h_{body}$. The final stage is achieved when the projectile is fully submerged, $S\geq h_{cone}+h_{body}$.

For convenience, the nonlinear equation of motion is expressed in the familiar aeroelasticity form:
\begin{equation}\label{eq:eomcd}
    (M+m) \ddot{S}-\frac{1}{2} \rho C_d(S) A_b(S) \dot{S}^2-  \rho g V_s(S)  = -Mg
\end{equation}
The second and third terms are often referred to as the hydrodynamic damping and hydrodynamic stiffness, respectively \citep{bisplinghoff2013principles}.  

\subsubsection{Added mass before and after reaching peak force impact}
The total added mass can be calculated before and after the peak impact events by integrating $dm/dS$ over the penetration depth.  
\begin{equation}
    m =m_{p^-}+m_{p^+}= \int_{0}^{L_p} \frac{dm_{p^-}}{dS} dS\ + \int_{L_p}^S \frac{dm_{p^+}}{dS} dS\\
\end{equation}
where $m_{p-}$ and $m_{p+}$ are the added mass before and after peak impact force. The added mass before reaching the peak impact can be approximated using the well-established \cite{shiffman1951force} relation
\begin{align}\label{eq:m}
    m_{p^-} = \rho k (d/2)^3 = \rho k \left(\tan{\beta}\right)^3S^3 = \rho \kappa S^3
\end{align}
where, $k$ is the dimensionless added mass coefficient. Moreover, we introduce the parameter $\kappa$, defined as $\kappa = k \tanh(\beta)$, to consolidate all geometric variables into a single parameter. Taking the derivative of the added mass before reaching peak force with respect to the penetration depth yields
\begin{align}\label{eq:dm}
    \frac{dm_{p^-}}{dS} = 3\rho k\left(\tan{\beta}\right)^3S^2 = 3 \rho \kappa S^2
\end{align}

The slope $dm/dS$ captures the cubic displacement of the added mass (\autoref{fig:ImpactForces}). Note that the peak force does not occur when the cone is fully submerged, as noted by\cite{vincent2018dynamics}. The change in added mass decays asymptotic and is modeled as an exponential function 
\begin{align}\label{eq:dm+}
    \frac{dm_{p^+}}{dS} &= 3\rho \kappa  L_p^2 e^{-\tau\sqrt{\frac{S}{Lp}-1}}
\end{align}
here $\tau$ is a parameter that indicates how fast added mass equalizes to its steady state value. The added mass portion after peak acceleration becomes
\begin{equation}
    m_{p+} = \frac{6}{\tau^2} \rho \kappa L_p^3  \left( 1-e^{-\tau \sqrt{\frac{S}{Lp}-1}} \left(\tau \sqrt{\frac{S}{L_p}-1}+1\right)\right)
\end{equation}

Expressing the added mass as a function of depth instead of time provides a mathematical convenience for solving the equation of motion. Therefore, the added mass can be treated in terms of the node given geometry at a specific depth. Therefore, the total added mass for $S\geq L_p$ becomes 
\begin{align}
    m_+ &= \rho \kappa L_p^3 + \int_{L_p}^S \frac{dm_{p+}}{dS} dS\\
    &= \rho\kappa L_p^3 + \frac{6}{\tau^2} \rho \kappa L_p^3  \left( 1-e^{-\tau \sqrt{\frac{S}{Lp}-1}} \left(\tau \sqrt{\frac{S}{L_p}-1}+1\right)\right)
\end{align}
To find all the unknown parameters — $\kappa$, $\tau$, and $L_p$ — a regression is conducted on all the collected rigid experimental data to determine these parameters. More details are provided in \autoref{appA}.

\begin{figure}
    \centering
    \includegraphics[width= \linewidth]{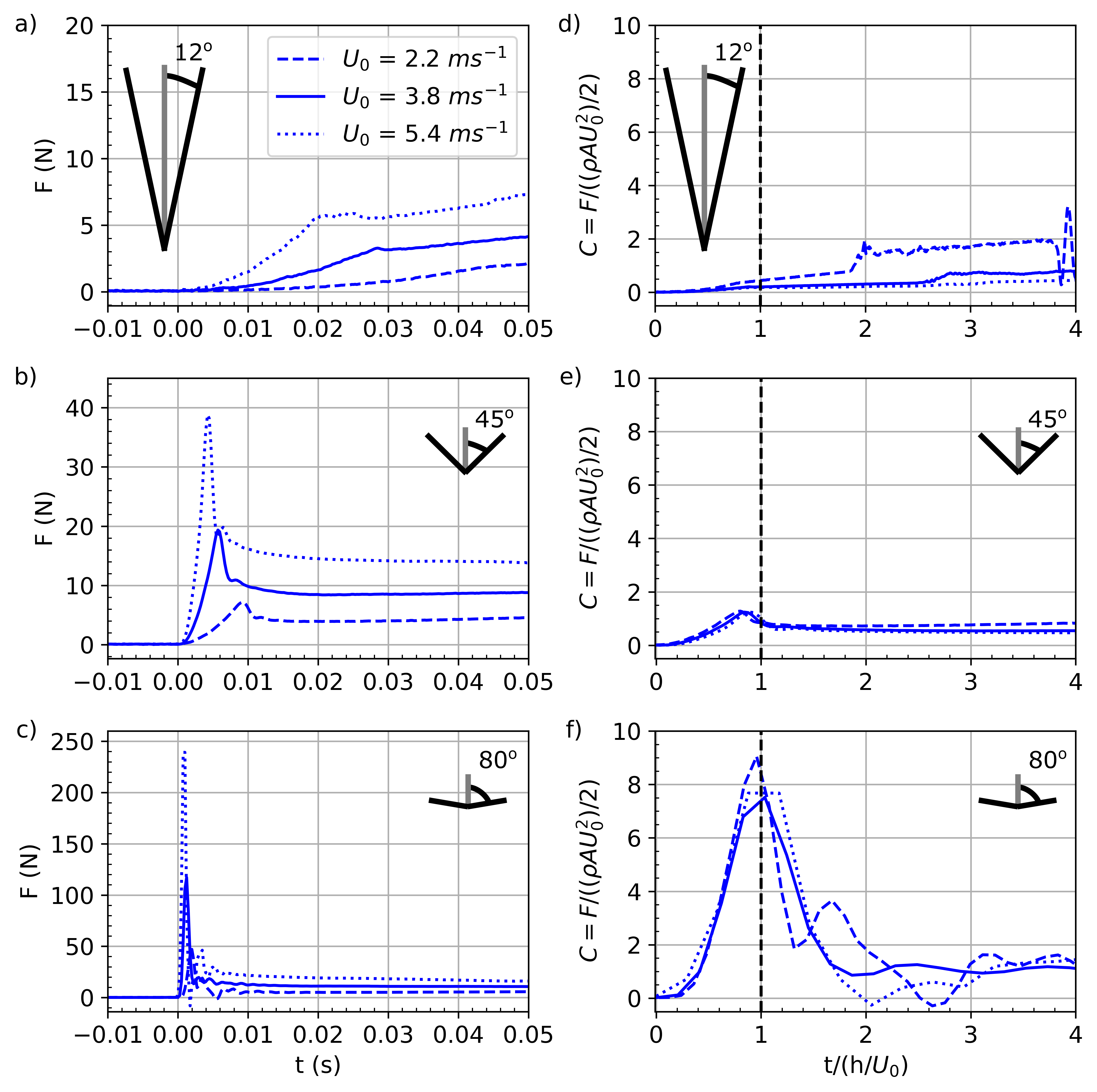}
    \caption{(a-c) dimensional and (d-f) non-dimensional water entry forces for noses with half angles $\beta=$ 12\textdegree, 45\textdegree\ and 80\textdegree\ in (a,d), (b,e) and (c,f), respectively. Note the y-axis scale is not the same for (a-c). The normalized lines(e-f) collapse into one line except the lower impact at half angle 12\textdegree\ (d). This is because at low speeds the buoyancy forces are more dominant for the smaller angles.}
    \label{fig:RigidData}
\end{figure}

\subsection{Result: effect of cone angles on impact forces}

The experimental data of the rigid projectiles with half angles $\beta = [12, 45, 80 ]$\textdegree\ is plotted as force versus time in \autoref{fig:RigidData} a-c, respectively. The forces were estimated by multiplying the acceleration measurements by the projectile mass, $M$. \autoref{fig:RigidData} (d-f) shows the non-dimensional force coefficient $C = F/(\rho A U_0^2/2)$ \citep{vincent2018dynamics} over non-dimensional time,  $t^*=t/(h_{cone}/U_0)$, where the base area is $A=\pi (d/2)^2$. The impact force is normalized by $(\rho A U_0^2)/2$, which is typically done when inertial forces are dominant. The time is normalized by $(h_{cone}/U_0)$ to obtain non-dimensional time $t^*$, which is the ratio of cone submersion when the cone is fully submerged at $t^* = 1$. 

From the non-normalized measurements, it can be observed that the projectile peak response force increases as $\beta$ and $U_0$ increases. For example, note the significant spike in the peak response force for a $\beta= 80$\textdegree\ nose compared to $\beta = 12$\textdegree, when holding the initial velocity constant. Hence, the water entry of a projectile with $\beta = 12$\textdegree\ is much smoother than at $\beta= 80$\textdegree. 

It can be noticed that in \autoref{fig:RigidData} d-f, the different impact velocities lay on top of each other when the inertia forces are dominant as in \autoref{fig:RigidData} e and f. Note that this is not the case at $\beta = 12$\textdegree, here the hydro-static forces like buoyancy are more prevalent and see a discrepancy in the non-dimensional force coefficient, see \autoref{fig:RigidData} d. This will also be the case when the mass of the impacting body is low. This will be the case when the body and head are separated. 

\begin{figure}
    \centering
    \includegraphics[width =\linewidth]{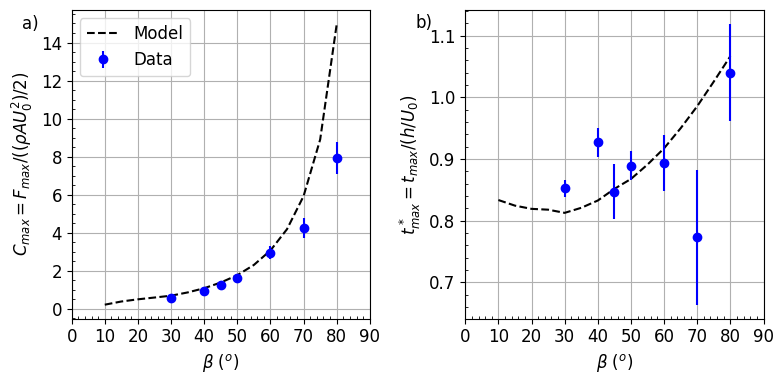}
    \caption{Comparison between the model and the experimental results for; a) the maximum drag coefficient $C_{max}=F_{max}/((\rho A U_0^2)/2)$, and the non-dimentionalized time at peak acceleration $t^*_{max}=t_{max}/(h/U_0)$.}
    \label{fig:CmaxRigid}
\end{figure}

The different nose angles are compared using two non-dimensional parameters, $C_{max}$ and $t^*_{max}$, which are the point maximal force coefficient and normalized time, respectively. These are graphically depicted in \autoref{fig:CmaxRigid} and numerically in \autoref{tab:ComparisonDataModel}. Since the lower angels do not exhibit any slamming forces, the force keeps increasing until it reaches a steady state; hence, these experimental peak forces in \autoref{fig:CmaxRigid} are reported for $\beta =30$\textdegree\ to 80\textdegree. The model however is able to capture where the transition of the added mass is, this point is taken for the angles 10 \textdegree\ to 25 \textdegree. For a more detailed comparison of the experimental data and the model see \autoref{tab:ComparisonDataModel}.  

\begin{table}
    \centering
    \begin{tabular}{c|cccc}
        Half angle $\beta$ & $C_{max}$ & $Cd_{max}$ (Model) & $t^*_{max}$ & $t^*_{max}$ (Model) \\
        10 & - & 0.207 & - &  0.833\\
        12 & - & 0.375 & - & 0.829\\
        25 & - & 0.584 & - & 0.818\\
        30 & 0.534$\pm$0.022 & 0.680 &  0.853$\pm$0.014& 0.812\\
        40 & 0.915$\pm$0.022 & 1.071 &  0.927$\pm$0.024& 0.831 \\
        45 & 1.263$\pm$0.033 & 1.364 &  0.847$\pm$0.045& 0.847\\
        50 & 1.585$\pm$0.118 & 1.755 &  0.889$\pm$ 0.024& 0.866\\
        60 & 2.934$\pm$0.340 & 3.049 & 0.893$\pm$0.045 & 0.917\\
        70 & 4.251$\pm$0.522 & 5.906 & 0.773$\pm$0.109 & 0.984\\
        80 & 7.926$\pm$0.823& 14.919 & 1.040$\pm$0.078 & 1.068
    \end{tabular}
    \caption{A Comparison between the experimental and modeling results of the drag and non-dimensional time at peak acceleration for various cone angles, $\beta$.}
    \label{tab:ComparisonDataModel}
\end{table}

\section{Spring stiffness effect on water-entry}\label{sec:stiffenss}
This section details the modeling development for estimating the water impact forces of segmented bodies connected by springs using nonlinear ordinary differential equations (ODEs). The objective is to compare the influence of the spring stiffness, and cone angle on the response forces. The model estimates of these forces are in good agreement with the experimental results.

\subsection{Estimating impact forces}

A diving system with a single array of $N$-segments (masses) connected by springs can be expressed as an $N$ degrees of freedom damped mass-spring system. As such, each segment $n$ and $n+1$ are connected by a spring, $k_n$, and viscous damper, $c_n$. For the sake of completeness, the matrix form of $N$-degrees of freedom system is provided in \autoref{appA}, which is the topic of future studies. This study focuses on a N=2 system, as described above. Additional information can be found in most vibration textbooks \citep{bhat2018principles}. The general form of the second-order ODE is:
\begin{equation}\label{eq:GenEoM}
    [\boldsymbol{M}] {\boldsymbol{\ddot{q}}}+ [\boldsymbol{C}] {\boldsymbol{\dot{q}}}+[\boldsymbol{K}] {\boldsymbol{{q}}}+ {\boldsymbol{{\mathcal{F}_{nl}}(q,\dot q)}}= {\boldsymbol{Q}}
\end{equation}

where, $\boldsymbol{q}=[x_1, S]^T$ is the displacement vector of the segments. The vector $\boldsymbol{Q}$ contains the gravitational forces. The hydrodynamic forces are captured in the lumped nonlinear force vector $\boldsymbol{{\mathcal{F}_{nl}}}$. These forces are the consequence of the time-variant unsteady velocity of the projectile. The matrices $[\boldsymbol{M}]$, $[\boldsymbol{C}]$, and $[\boldsymbol{K}]$ are the system mass, damping, and stiffness components.  In this investigation, the projectile prototypes (\autoref{fig:ExpermintalSetup}-b) were one-dimensional damped two-mass spring systems. Therefore, the model includes the effect of the vertical oscillations of the system on the instantaneous hydrodynamic forces, which depend not only on the instantaneous depth position, $S(t)$ of the nose ($m_2$), but also the time-varying nose velocity, $U(t)=\dot{S}$. To this end, the system's mass, damping, and stiffness matrices and force vector can be expressed as follows

\begin{align*} \label{eq:Q}
    \boldsymbol{M} = &\begin{bmatrix} m_1 & 0 \\ 0 & m_2 \end{bmatrix}, &&\boldsymbol{C} = \begin{bmatrix} c & -c \\ -c & c  \end{bmatrix}, 
    &&\boldsymbol{K} = \begin{bmatrix} k & -k \\ -k & k \end{bmatrix},\\\boldsymbol{Q} =& \begin{bmatrix} F_g\\ F_g \end{bmatrix},
    &&\boldsymbol{\mathcal{F}_{nl}} =\begin{bmatrix} 0\\ F_{Cd}+F_b \end{bmatrix}
\end{align*}

To simplify the discussion of the underlying physics in the equation of motion, $\mathcal{F}_{nl}$ nonlinear inertia, hydrodynamic damping, and hydrodynamic stiffness coefficients are combined with $[\boldsymbol{M}]$, $[\boldsymbol{C}]$, and $[\boldsymbol{K}]$, respectively. Accordingly, \autoref{eq:GenEoM} equation is rewritten to group all the coefficients related to $q$, $\dot{q}$ and $\Ddot{q}$

\begin{equation} \label{eq:HydroEoM}
    [\Tilde{\boldsymbol{M}}]\boldsymbol{\ddot{q}}+
    [\Tilde{\boldsymbol{C}}]{\boldsymbol{\dot{q}}}+[\Tilde{\boldsymbol{K}}]{\boldsymbol{{q}}}= {\boldsymbol{[{\boldsymbol{Q}}]}}
\end{equation}

The observable mass, damping, and stiffness matrices include the linear and nonlinear physics of the system, which can be expressed, respectively, as
\begin{align}
    \Tilde{\boldsymbol{M}} &= [\boldsymbol{M}] + [\boldsymbol{B}] ,
    &\Tilde{\boldsymbol{C}} = [\boldsymbol{C}] -\frac{1}{2} \rho [\boldsymbol{D}] {\boldsymbol{\dot{q}^2}}, \qquad
    &\Tilde{\boldsymbol{K}} = [\boldsymbol{K}] -\rho g [\boldsymbol{K_b}]
\end{align}
where,
\begin{align}
\boldsymbol{B} &= \begin{bmatrix} 0 & 0 \\ 0 & m \end{bmatrix}, 
&\boldsymbol{D} = \begin{bmatrix} 0 & 0\\ 0 & C_{d}(S)A_b(S) \end{bmatrix}, \qquad
&\boldsymbol{K_b} = \begin{bmatrix} 0 & 0 \\ 0 & V_s(S) \end{bmatrix}
\end{align}

The nonlinear terms are the consequence of the hydrodynamics physics of the system. The matrix $\boldsymbol{B}$ is the hydrodynamic inertia of the added mass. The hydrodynamic damping and hydrodynamic stiffness matrices are $\boldsymbol{D}$, and $\boldsymbol{K_b}$, respectively. It can be noticed that $\boldsymbol{B}$, $\boldsymbol{D}$, and $\boldsymbol{K_b}$ are response-dependent matrices. That is, they change as a function of the projectile's depth and velocity. Here the components of these matrices can be found in \autoref{sec:shape}. To this end, the new arrangement of \autoref{eq:GenEoM} illustrates how the nonlinear hydrodynamic terms inertia, damping, and stiffness are coupled with the projectile's linear mass, viscous damping, and spring terms, respectively. Therefore, the system's apparent mass, damping, and stiffness are $[\boldsymbol{M}] +\rho [\boldsymbol{B}] q^3$, $[\boldsymbol{C}] -\rho/2 [\boldsymbol{D}] {\boldsymbol{\dot{q}^2}}$, and $[\boldsymbol{K}] - \rho g [\boldsymbol{K_b}]$, respectively. The term $[\boldsymbol{B}]$ can be thought of as the inertial force of the displaced water over time. The hydrodynamic damping force $\rho/2 [\boldsymbol{D}]  {\boldsymbol{\dot{q}^2}}$ is a quadratically dissipative term with quadratic velocity term due to drag. The buoyancy force $\rho g [\boldsymbol{K_b}]$ acts as spring restoring force but behaves cubically and linearly since it is a function of the projectile volume--- note $V_s$ is a cubic function of the depth, \autoref{eq:C_d}.  

The importance of \autoref{eq:HydroEoM} formulation is that it provides an accurate estimation of the projectile dynamics since it captures the temporal evolution of the hydrodynamic mass, damping, and stiffness matrices, starting from impact initiation. Therefore, the hydrodynamic coefficients must be updated for every time step. This is because during an unsteady condition, as the projectile velocity changes, and in term the impact forces change. In \autoref{fig:SpringModel} the hydrodynamic damping and stiffness forces ramp-up while oscillating over time, followed by fluctuating decay. For example, the buoyancy force oscillates due to harmonic compression and stretching in the projectile volume in response to the spring oscillations. Similarly, the drag force experiences decay with periodic fluctuation due to harmonic velocity response between compression and extension. The added mass experiences similar dynamics. Detailed discussion on oscillatory behaviors from an energy perspective is provided in \autoref{sec:energy}.

\subsection{Results: stiffness effect of sprung segment on impact forces} \label{sec:segmentedDiver}
\begin{figure}
    \centering
    \includegraphics[width =\linewidth]{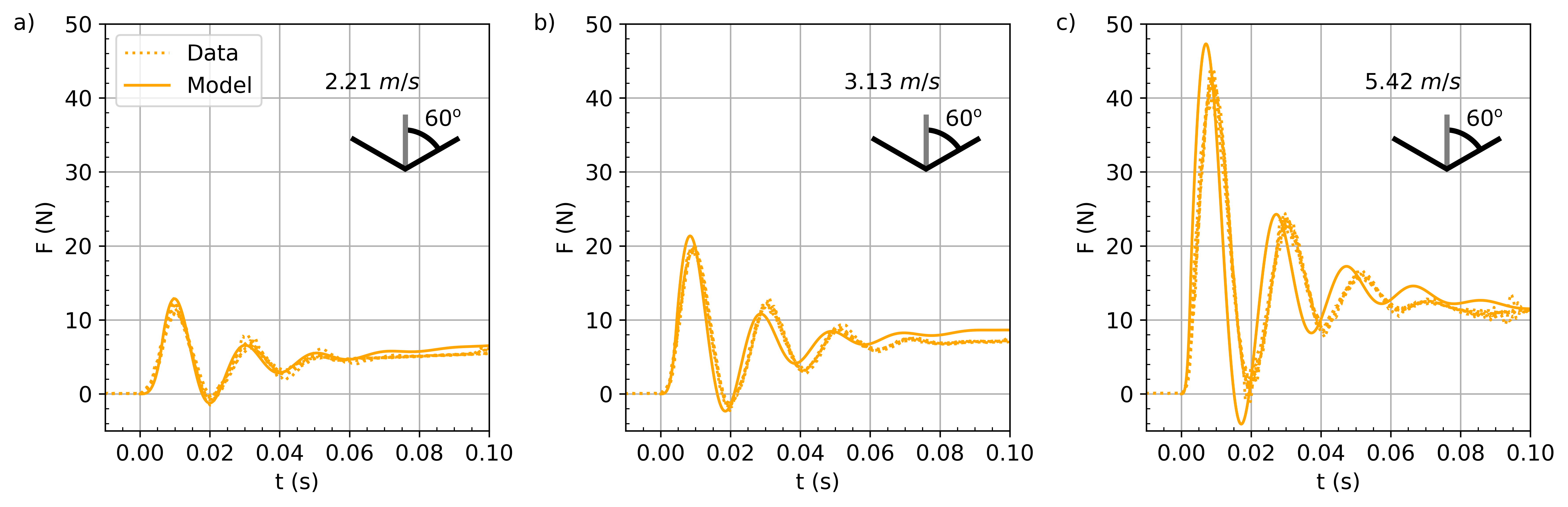}
    \caption{Alignment of the modeling and experimental results for a firm spring at different impact velocities for a cone angle of 60\textdegree and the firm spring (7.8 N/mm).}
    \label{fig:SpringModel}
\end{figure}

The results of the analytical model, \autoref{eq:HydroEoM}, are in good agreement with the experimental results; see, for example, \autoref{fig:SpringModel}. This figure compares the impact forces obtained from the model to the experimental results for a projectile with a firm spring and 60\textdegree\ cone angle at velocities 2.2, 3.1, and 5.4 m/s. For this projectile, a damping coefficient of $c$ = 10 is chosen, thus, the damping ratio is $\zeta = c/2\sqrt{km_2(1-\frac{m2}{m_1+m_2})} \approx 0.2$. There is a minor deviation in the peaks when comparing the model to the experimental results, which is attributed to a slightly overestimated stiffness. Nonetheless, the trend of the dynamics over time is within a reasonable agreement.

\begin{figure}
    \centering
    \includegraphics[width=\linewidth]{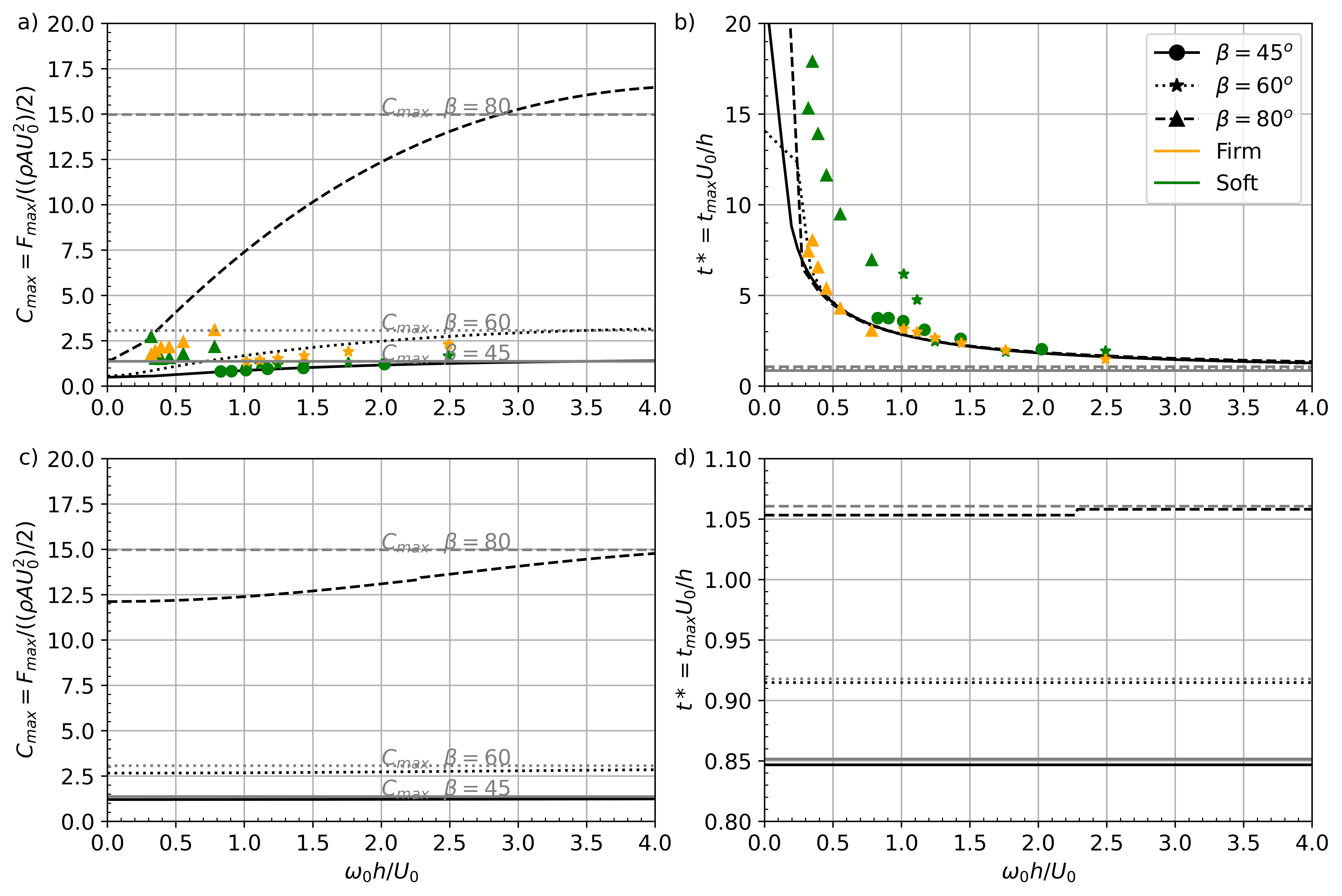}
    \caption{a,c) Show the $C_{max}$ for different cone angles of the body and the head, respectively. Here, they are compared to the $C_{max}$ of the rigid model shown as horizontal lines due to no change in maximal drag force over impact velocity. b,d) Show the normalized time to maximal impact over normalized eigenfrequency. Note that there is only one model line since the normalized time to impact scales directly to normalized eigenfrequency.} 
    \label{fig:BodyCdmaxtmax}
\end{figure}

\autoref{fig:BodyCdmaxtmax} shows the maximal force coefficient $C_{max}$ and normalized time at peak impact $t^*_{max}$ as a function of the normalized eigenfrequency $\omega_0 h / U_0$. Contrary to the rigid projectile the force coefficient is not constant over the impact velocity. This is due to the internal dynamics of the entering body since the spring system phase-gain response is a function of the excitation frequency, i.e. the impact velocity. This is also different for the head and the projectile's body since they experience different forces and have different masses.

\autoref{fig:BodyCdmaxtmax} a) shows the evolution of the maximal force coefficient as a function of the non-dimensional eigenfrequency. In this study, the eigenfrequency is normalized with $h/U_0$, which is approximately the impulse frequency. It can be observed that the maximal force coefficient is smaller for low impact velocity or soft springs. Similarly in \autoref{fig:BodyCdmaxtmax} b), it can be observed that the $t^*_{max}$ increases with a decreasing eigenfrequency or increasing impact velocity. \autoref{fig:BodyCdmaxtmax} c) and d) show that the $t^*_{max}$ is similar and only slightly experiences peak force before the rigid case. However, the shape still has the dominant effect on the maximal force coefficient, see \autoref{fig:BodyCdmaxtmax} -c. Although a similar trend, the magnitude of the response differs greatly. The nose force coefficient is lower for low $\omega h/ U_0$ compared to the rigid case. It can be observed that the effect of a higher nose angle is greater than the sharp noses. Here the reduction in maximal velocity is greatest and thus the maximal force to the head. The force coefficients of the head and the body are slightly differently defined. For the body, it is calculated as the maximal acceleration times the mass of the body, expressed as $C = m_1 \ddot{x}_1 / ((\rho A U_0^2)/2)$. The head is defined differently due to forces acting on it from two directions: impact forces from below and the spring and damper from above. Therefore, the force coefficient is calculated as follows
\begin{equation}
    C = (m_2\Ddot{x}_2 - k(x_2-x_1) - c(\dot{x}_2-\dot{x}_1))/ ((\rho A U_0^2)/2)
\end{equation}
This formulation determines the change in the effect of the impact forces on the head.

\section{Performance of shape vs. sprung bodies}\label{sec:Results}
This section discusses the influence of the interplay between the projectile's spring stiffness and nose angle on water entry performance. In general, the results show that integrating springs into a projectile may provide passive phase (time lag) control of the non-conservative energy due to drag, i.e., delay the energy losses through elastic storage in the springs. This conclusion is supported by the experimental and analytical results reported in this section, which illustrate the effect of the stiffness and head shape interplay the energy flow, efficiency of the impact, and body jerk.

\subsection{Energy flow in the system} \label{sec:energy}
\begin{figure}
    \centering
    \includegraphics[width = \linewidth]{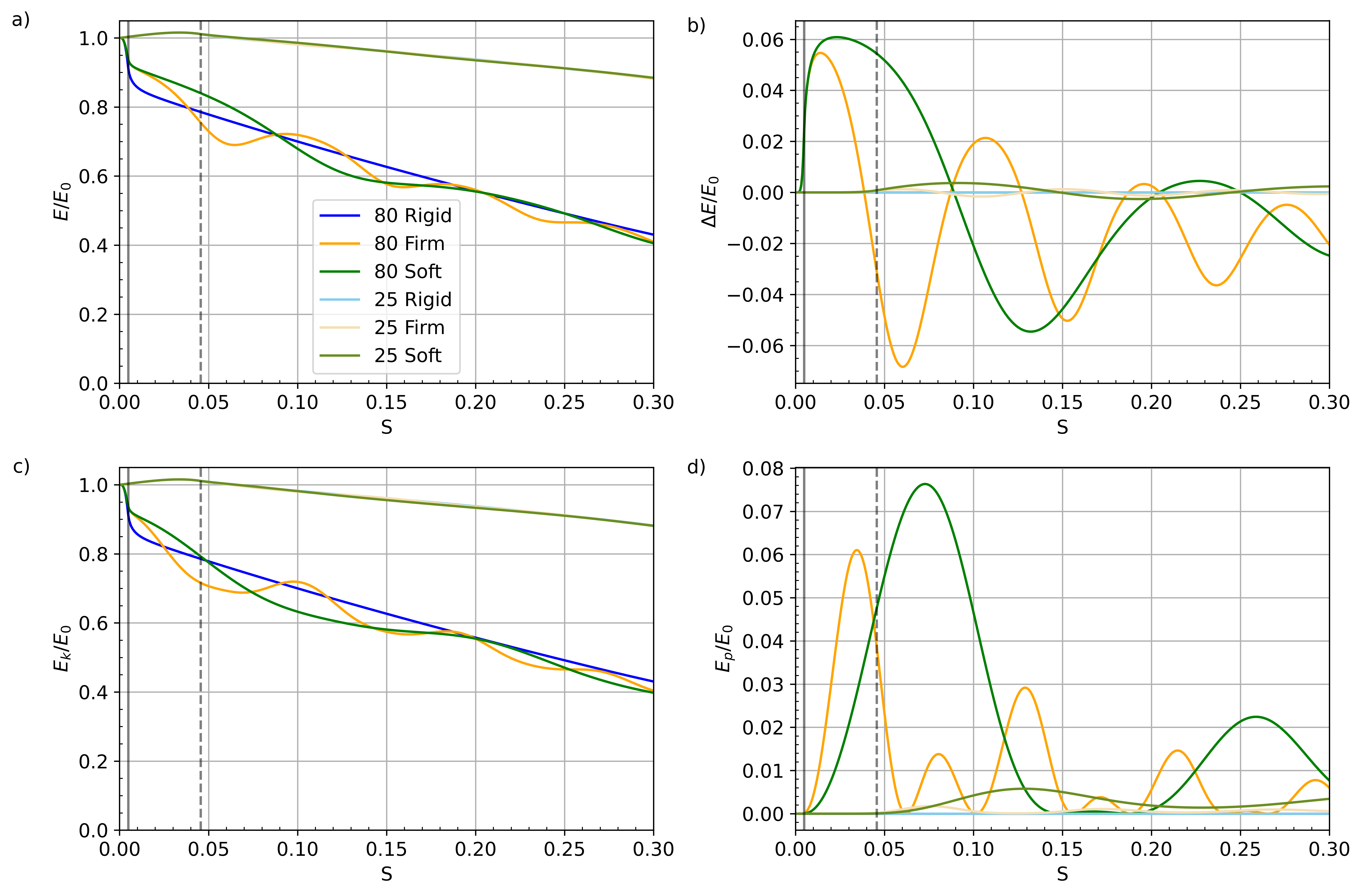}
    \caption{The energy in all plots is normalized by the initial energy $E_0$, which is the energy at $S=0$. The solid and dashed grey vertical lines are the $L_p$ of 80\textdegree\ and 25 \textdegree\, respectively. a) shows the total energy in the system from the moment the nose touches the water until the point where the whole body is submerged. b) shows the energy difference between the rigid and the sprung cases, here a positive value means that there is more energy in the sprung system compared to its rigid counterpart and vice versa. c) shows the kinetic energy. d) shows the potential energy stored in the spring.}
    \label{fig:EnergyLoss}
\end{figure}

The model developed in \autoref{sec:forceModel} was utilized to calculate the energy flow and distribution in the system. The energy flow is defined as the change in the energy over the water penetration depth. The energy experienced by the system consists of: 

\begin{align}
     E_0 &= \frac{1}{2}mU_0^2, \quad \text{where} \quad m = m_1 + m_2\\
     E_{Sprung} &= E_{k} + E_{p} = \frac{1}{2}m_1\dot{x_1}^2 + \frac{1}{2}m_2\dot{S}^2  + \frac{1}{2}k(x_1-S)^2\\
     E_{Rigid} &= \frac{1}{2}m\dot{S}^2\\
     \Delta E &= E_{Sprung}-E_{Rigid}
\end{align}

where, $E_0$, $E_{Sprung}$, $E_{Rigid}$, are the initial energy at impact, $t^+=0$, and total energy for the sprung and rigid systems, respectively. To compare the energy of the rigid and segmented projectiles, $\Delta E$ is calculated as the difference between $E_{rigid}$, $E_{sprung}$. $E_{Sprung}$ is expressed as the sum of the kinetic energy due to the projectile inertia and the potential energy of the spring required to restore the system back to equilibrium. It was assumed that the conservative potential energy due to buoyancy for the rigid and the sprung cases were equal. Therefore, the buoyancy potential energy was omitted from the energy flow calculations. \autoref{fig:EnergyLoss} illustrates the energy flow in the system as a function of depth. The energy values are normalized by $E_0$ and provided for 25\textdegree\ and 80\textdegree\ cases with and without springs. These two extreme nose angles were chosen to provide a succinct discussion of the analysis for efficient and worst angles for water entry, which were 25\textdegree\  and 80\textdegree\, respectively. 

The energy shown in \autoref{fig:EnergyLoss}-a is calculated from depth $S=0$, where the nose touched the water surface, to $S=0.3$ until the whole body was submerged. From $S=0$ to $S=L_p$, depth at peak acceleration, it can be observed that the energy of a projectile with $\beta = 80$\textdegree\ dissipates significantly faster. The drops in the total energy are $15\%$ and $7\%$ for the rigid and sprung projectiles at peak force, respectively. However, projectiles with the 25\textdegree\ head continue to accelerate for the rigid and sprung systems at initial impact. This is because the gravitational force is higher than the drag forces. After $S=L_p$ is reached, the total energy decreases linearly for the rigid projectiles regardless of head angle size, see \autoref{fig:EnergyLoss}-a. It can be noticed that for sprung systems with large $\beta$, however, the total energy drops fluctuate harmonically when the projectiles exceed the peak acceleration depth, $S>L_p$. These fluctuations are compared with the rigid cases by plotting $\Delta E$ versus the depth, $S$, and normalized by the input energy $E_0$, as shown in \autoref{fig:EnergyLoss}-b. A positive $\Delta E$ means the sprung system has more energy than the rigid one at the same depth, and vice versa. The energy fluctuation amplitudes for the sprung projectile with 80\textdegree\ nose are significantly higher than those with 25\textdegree\ nose regardless of the spring stiffness. Thus, introducing springs in a small nose projectile does not appear to have a significant effect on the energy flow. The maximum $\Delta E$ for 80\textdegree\ with firm and soft springs are 15\% and 13\% higher than those for 25\textdegree, respectively (\autoref{fig:EnergyLoss}-b). It can be concluded that projectiles with narrow nose angles are closer to the performance of rigid systems-- hence the low amplitude fluctuations in the energy of the 25\textdegree\ projectiles. The fluctuations are higher for the $\beta = 80$\textdegree\ due to the high impulse force, see \autoref{fig:RigidData}; therefore the spring can absorb more energy.

\autoref{fig:EnergyLoss}-c and -d show the kinetic and potential energy evolution, respectively, as a function of depth for rigid and sprung projectiles with 25\textdegree\ and 80\textdegree\ nose angles. The potential energy can be interpreted as the stored energy in the spring that can be absorbed from the kinetic energy. One of the most important observations in the energy analysis is the asymmetric fluctuations in both the kinetic and potential energy of all sprung systems. The asymmetric decay is attributed to the simultaneous energy transfer between kinetic and potential while experiencing energy losses due to drag.  When the spring is compressed the body length shortens as well. Subsequently, the body's absolute velocity becomes lower than that of the rigid case, which reduces the drag forces. Conversely, when the spring is extended, the body stretches, which causes an increase in the drag forces. This explains the delay in energy losses due to drag. Therefore, it is incorrect to assume constant velocity during water entry for segmented bodies. Assuming constant velocity implies that the drag force is also constant, which is inaccurate for high acceleration non-steady states. For the sake of completion, it is worth mentioning that internal mechanical friction forces between the head and the body existed and increased energy losses. Thus, the energy losses are attributed to internal friction and drag forces. To this end, it was necessary to estimate the efficiency of an impulse to approximate the energy losses in the system, which is discussed in \autoref{sec:impulse}.

\begin{figure}
    \centering
    \includegraphics[width=0.5\linewidth]{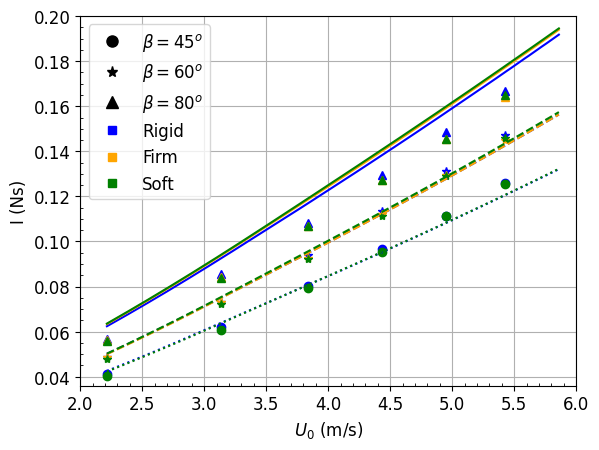}
    \caption{Impulse of water entry as a function of the initial velocity for different nose angles and spring stiffnesses. The blue, orange, and green colors represent the rigid, firm, and soft springs, respectively. The solid, dashed, and dotted lines represent the model for half angle $\beta$= 80\textdegree\, 60\textdegree\, and 45\textdegree\, respectively. The triangle, star, and circle markers are calculated from the experiments for half angle $\beta$= 80\textdegree\, 60\textdegree\, and 45\textdegree, respectively.}
    \label{fig:Impulse}
\end{figure}

\subsection{Impulse} \label{sec:impulse}
The efficiency of an impact, also known as the impulse, $I$, is analyzed for projectiles with three different half angles, $\beta = [45\text{\textdegree},  60\text{\textdegree}, 80\text{\textdegree}]$, and three stiffness levels, rigid, firm, and soft. The impulse, $I$, was quantified by evaluating the change in momentum over the impact duration, $t$. The impulse is considered an adequate approximation of the energy losses in the system due to impact. The acceleration measurements were utilized to calculate $I$ using the following integral:
\begin{equation}
    I =  \int_0^{t_s} F dt = \int_0^{t_s} m_1 \Ddot{x}_1 dt 
\end{equation}
where $t_s$ is set to 100 ms. This specific time duration is selected because the vibrations have attenuated beyond this point, and both the rigid and spring cases have reached a steady state. The resulting impulse values are depicted in \autoref{fig:Impulse}. It is evident from the experimental and modeling results in the figure that the impulses experienced by the sprung projectiles are consistently lower than those for the rigid case. This insignificant finding indicates that integrating spring elements into a projectile does not alter the total impact efficiency. The impulse looks to be weakly quadratic; as expected, the blunter nose angles show a bigger impulse. 

\subsection{Jerk}\label{sec:jerk}

Jerk is another important analysis that was performed in this study as an indicator of projectile survivability. A high jerk often excites high-frequency dynamics that can damage the body experiencing instantaneous changes in impact acceleration. This is why jerk, $J$, is a widely used impact severity indicator in automotive crash testing and orbital dynamics control to meet safety and survivability requirements \citep{eager2016beyond}. To conduct an adequate comparison analysis for the various projectile designs in this study, we employed the normalized $J^*$ introduced by \cite{sharker2019water}
\begin{equation}
    J^* = \frac{\Delta a}{\Delta t} \frac{m}{(\rho g A U_0^2)/2}
\end{equation}
where $\Delta a$ and $\Delta t$ are defined linearly from $t^+=0$ to the point of peak acceleration, where $S=L_p$. 

\begin{figure}
    \centering
    \includegraphics[width =\linewidth]{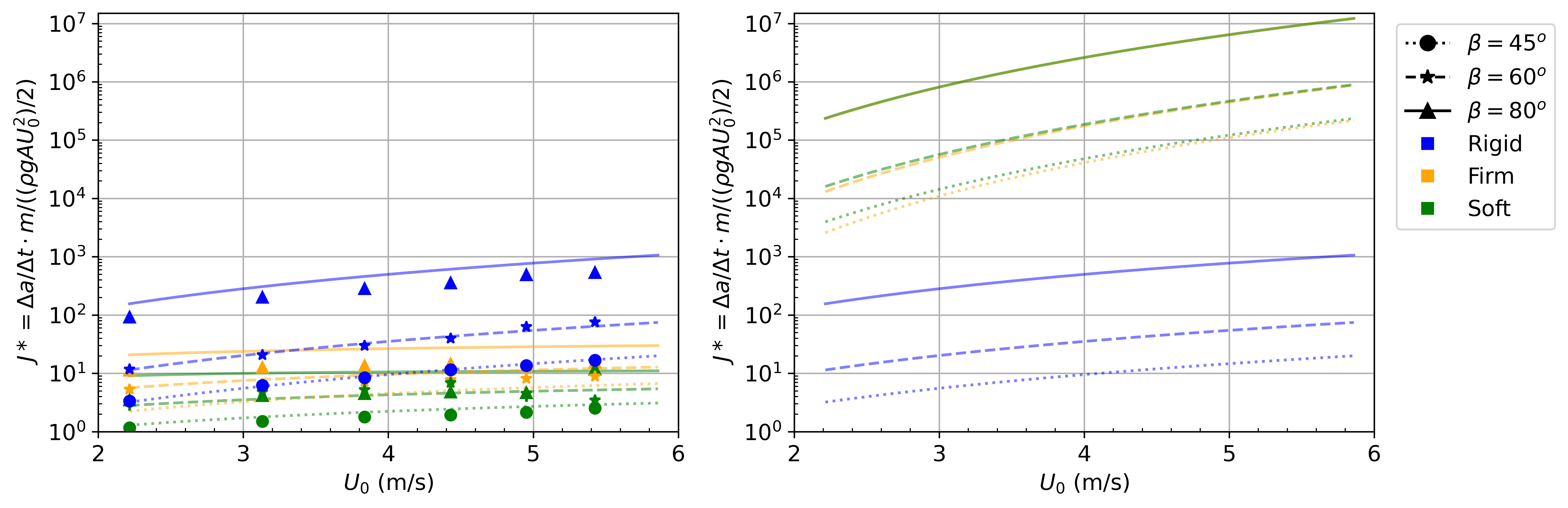}
    \caption{Comparing normalized jerk over impact velocity for different nose angles and stiffness. a) Shows the jerk of the body and b) the jerk of the nose. The colors blue, orange, and green represent the rigid, firm, and soft springs, respectively. The solid, dashed, and dotted lines represent the model for half angle $\beta$= 80\textdegree\, 60\textdegree\, and 45\textdegree\, respectively. The triangle, star, and circle markers represent the experimental data for half angle $\beta$= 80\textdegree\, 60\textdegree\, and 45\textdegree\, respectively. }
    \label{fig:Jerk}
\end{figure}

The jerk can be thought of as the rate of change in the forces acting on a body. 
Reducing the maximal jerk has the advantage of stretching the time for the system to react to the input force. \autoref{fig:Jerk}-a and -b show the normalized jerk for the body and nose, respectively. The curves indicate the modeling results, where blue, orange, and green represent the rigid, firm, and soft stiffnesses, respectively. Additionally, the dotted, dashed, and solid curves represent the half nose angle $\beta = [45\text{\textdegree}, 60\text{\textdegree}, 80\text{\textdegree}]$, respectively. The markers in \autoref{fig:Jerk}-a are the experimental results. In both \autoref{fig:Jerk}-a and -b it can be noticed that the impact velocity influences the jerk regardless of the nose shape or stiffness. However, \autoref{fig:Jerk}-a illustrates how narrow noses and springs can drastically reduce the jerk to the body. Notice that the slope of the sprung cases is much smaller than the rigid cases. For the rigid case, the difference between the 80\textdegree\ and 45\textdegree\ noses can reduce the jerk by 98\% at $U_0 = 5.4$ m/s. Remarkably, a similar reduction of 99\% can be obtained when a soft spring is integrated into an 80\textdegree\ projectile for the same velocity. Even with a firm spring, $4.5\times k_S$, the jerk is reduced by approximately 97\%. 

The nose jerk exhibits different behaviors than that of the body. The jerk of the nose is increased by introducing a spring, as depicted in \autoref{fig:Jerk}-b. This is expected since the head mass is significantly lower than the rigid body; for example, the 80\textdegree\ nose mass is 1/10th of the rigid body. In other words, a lighter nose will accelerate faster than a rigid body. For the 80\textdegree\ nose, the head jerk in the sprung cases increases by four orders of magnitude compared to the rigid projectile at $U_0 = 5.4$ m/s. The jerk for projectiles with firm and soft springs are in the same order of magnitude.

\section{Conclusion}
We studied the combined effects of the aerodynamic shape of a nose cone and the stiffness of the spring connected to a rear body to mitigate slamming forces during water entry. Unlike current models, the hydrodynamic nonlinearities of the stiffness and damping were not ignored in our unsteady state analysis. This was a key to uncovering two major advantages of incorporating a coupled spring: (i) reducing body jerk across sprung cases up to $\approx$ 98\%, and (ii) decreasing impact peak forces, especially for diving systems with blunt cone angles by $\approx$ 90\%. These remarkable advantages were achieved with a minimal decline in water penetration efficiency, measured by impulse, with the worst-case scenario showing a 1.5\% difference in the impulse. The key mechanisms behind these benefits were attributed to periodic fluctuations between the kinetic and potential energy, tuned by spring stiffness to control the periodic energy storage and release. Inspired by the geometry of diving bird necks, we aim to investigate the influence of multiple coupling springs in segmented structures on their dynamic response during high-velocity water entry in future studies.

\appendix
\section{Model parameters estimation}\label{appA}
\begin{figure}
    \centering
    \includegraphics[width=\linewidth]{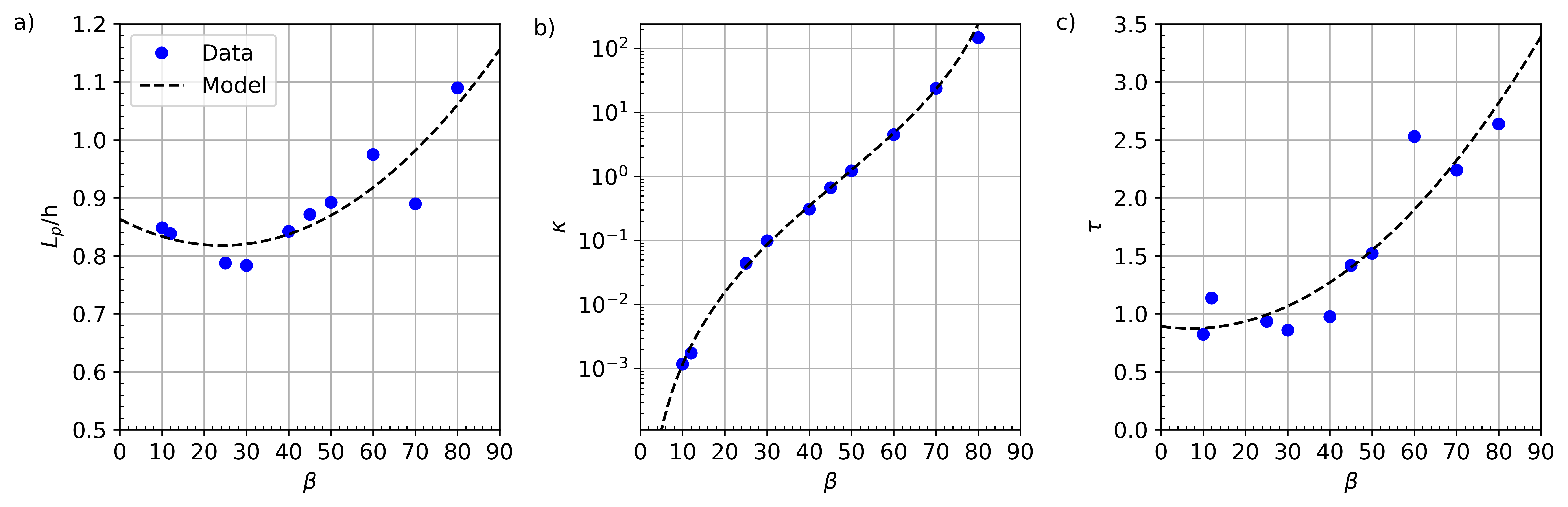}
    \caption{The model parameters estimations are obtained by applying least square fitting. The blue markers are the results from each optimization. The black dotted curves are the fitted estimations. a) shows the submerged cone depth at peak acceleration, b) the added mass geometric parameter, $\kappa$, and c) the decay variable $\tau$.}
    \label{fig:NewFitting}
\end{figure}

To estimate $C_{ds}$, $L_p/h$, $\kappa$, and $\tau$, least squares regression is utilized using simulated and experimental data. The continuous relationships for these parameters as a function of the cone angle, $\beta$, can be expressed as
\begin{align} \label{eq:fitFunctions}
    C_{ds} &= a \beta^3 + b \beta^2 + c\beta +d  \\
    L_p/h  &= a \beta^2 + b\beta +c\\
    \kappa &= (a \beta^2 + b\beta +c)(\tan \beta)^3\\
    \tau   &= a \beta^2 + b\beta +c
\end{align}

where, $a$, $b$, $c$, and $d$ are the fitting coefficients, which are given in \autoref{tab:FitParameters}. This results in a continuous model over the full range of cone angles. The steady state coefficient of drag $C_{ds}$ from \cite{baldwin1971vertical} is utilized and fitted using third-degree polynomial, as shown in \autoref{eq:fitFunctions} and \autoref{tab:FitParameters}. The least squares cost function is the difference between the model and simulated data of all trials for a specific cone shape. For example, the residuals between the simulations for all heights are compared with all data trials. The optimal estimates for $L_p$, $\kappa$, and $\tau$ are found by minimizing that residual. This leads to a set of parameters for every tested cone angle, as shown by the markers in \autoref{fig:NewFitting}. 

\begin{table}
    \centering
    \begin{tabular}{l|c|c|c|c}
        &  $C_{ds}$ & $L_p/h$ & $\kappa$ & $\tau$  \\
        a & $3.4\times 10^{-2}$ & $7.80\times 10^{-5}$ & $8.17\times 10^{-5}$  & $3.67\times 10^{-4}$ \\
        b & $-2.7\times 10^{-1}$ & $-3.77\times 10^{-3}$  & $8.38\times 10^{-3}$ & $-5.19\times 10^{-3}$\\
        c &  $8.5\times 10^{-1}$  & $8.63\times 10^{-1}$ & $1.21\times 10^{-1}$ & $8.92\times 10^{-1}$ \\
        d & $-3.8\times 10^{-2}$ & -  & -  & -\\
    \end{tabular}
    \caption{Fitting coefficients found by the least square optimization procedure to estimate the model parameters: $C_{ds}$, $L_p/h$, $\kappa$, and $\tau$.}
    \label{tab:FitParameters}
\end{table}

\section{n-degree of freedom dynamic model}\label{appB}
The model can be easily extended to multiple segments since only the head is in contact with the water surface. We can make the assumption that only the head experiences hydrodynamic forces. Therefore, the n-degrees of freedom (n-DoFs) model can be expressed as
\begin{equation}
\begin{split}
    \begin{bmatrix}
        m_1 & 0 & 0 & 0\\
        0 & m_2 & 0 & 0\\
        \vdots & \ddots & \ddots & \vdots \\
        0 & 0 & 0 & (m_n+m)
    \end{bmatrix}\begin{bmatrix}
        \Ddot{x}_1\\
        \ddot{x}_2\\
        \vdots \\
        \ddot{x}_n
    \end{bmatrix} 
    + \begin{bmatrix}
        c_1 & -c_1 & 0 & 0 \\ 
        -c_1 & c_1+c_2 & -c_2 & 0\\
        \vdots & \ddots & \ddots & \vdots \\
        0 & \hdots & -c_n & c_n \\
    \end{bmatrix}
    \begin{bmatrix}
        \dot{x}_1\\
        \dot{x}_2\\
        \vdots \\
        \dot{x}_n
    \end{bmatrix} +\\
    \begin{bmatrix}
        k_1 & -k_1 & 0 & 0 \\ 
        -k_1 & k_1+k_2 & -k_2 & 0\\
        \vdots & \ddots & \ddots & \vdots \\
        0 & \hdots & -k_n & k_n \\
    \end{bmatrix}\begin{bmatrix}
        x_1\\
        x_2 \\
        \vdots \\
        x_n
    \end{bmatrix} - \begin{bmatrix}
        m_1g \\ 
        m_2g \\
        \vdots \\
        m_ng
    \end{bmatrix} =  \begin{bmatrix}
        0 \\
        0 \\
        \vdots\\
        F(x_n,\dot{x}_n)
    \end{bmatrix}
    \end{split}
\end{equation}

\subsection{Eigenfrequency calculation}
The eigenfrequency can be utilized to normalize the results in \autoref{sec:Results}. Note that this is the damped resonant frequency since the damping and the nonlinear effects are neglected. Calculating the eigenfrequency of the two DoFs undamped homogeneous linear system can be achieved by omitting the nonlinear terms and damping. The procedure can be found in most vibration textbooks \citep{meirovitch2000fundamentals}.
\begin{align}
    det([K-\omega^2M])=0\\
    m_1m_2\omega^4 - km_2\omega^2 - km_1\omega^2 = 0 
\end{align}
The only acceptable solution is given by the following equation
\begin{equation}\label{eq:w0}
       \omega_0 = \frac{\sqrt{km_1m_2(m_1 + m_2)}}{m_1m_2}
\end{equation}

\bibliographystyle{jfm}
\bibliography{jfm-instructions}

\end{document}